\documentclass{kluwer}
\usepackage{graphicx}
\begin{document}
\begin{article}
\begin{opening}
\date{}

\title{GAMMA RAYS FROM  MOLECULAR CLOUDS}

\author{F.A.  AHARONIAN}
\institute{}
\institute{Max-Planck-Institut fu\"r Kernphysik, Heidelberg}

\runningauthor{F.A.~Aharonian}

\runningtitle{Gamma Rays from Molecular Clouds}

\begin{abstract}
High energy $\gamma$-rays from individual giant molecular 
clouds contain unique information about the  hidden sites 
of acceleration of galactic cosmic rays, and provide a feasible  
method for study of propagation of cosmic rays 
in the Galactic Disk on scales $\leq 100 \, \rm pc$.  
I discuss the spectral features of $\pi^0$-decay $\gamma$-radiation 
from clouds/targets located in proximity of relatively young 
proton accelerators,  and speculate that such ``accelerator+target'' 
systems in our Galaxy can be  responsible for a subset of unidentified  
EGRET sources. Also, I argue that the recent observations of high 
energy $\gamma$-rays from the Orion complex
contain evidence that the level of the ``sea'' of galactic cosmic rays
may differ significantly 
from the flux and the spectrum of local (directly detected) particles.  
\end{abstract}
\end{opening}

\section{Introduction}

The solution of the long-standing problem of the origin of galactic 
cosmic rays (GCRs)  depends, to a large extent, on advances  of 
$\gamma$-ray astronomy. The basic idea of this belief is straightforward 
and concerns both the acceleration  and propagation aspects of the problem -  
while the localized sources of $\gamma$-rays exhibit the {\em sites} of 
particle acceleration, the angular and spectral distributions of  
diffuse $\gamma$-radiation give information about the {\em propagation} 
of CRs in the Galactic Disk.  

The studies of the diffuse galactic  $\gamma$-radiation by EGRET aboard 
the Compton GRO have already made a significant contribution to the 
knowledge of  spatial distribution of low energy CRs in our 
Galaxy \cite{hunter}. Furthermore, many famous representatives of  
different classes of potential particle accelerators 
like  pulsars,  shell-type supernova remnants,
giant molecular clouds and  associated with them star formation 
regions, are identified  by EGRET as sources of 
$\geq 100 \, \rm MeV$ radiation \cite{egret3}. 
It should be noted, however, that 
most of the EGRET sources at low- and mid- galactic 
latitudes still do no have clear counterparts 
at other wavelengths \cite{Gehrels}. 
The next generation major space-based 
$\gamma$-ray detector, the GLAST  \cite{glast} 
with its superior flux  sensitivity (see Fig.~1) and good angular resolution 
will be able to reveal the nature of these hot spots.
GLAST offers  a broad - up to 100 GeV  
for bright and flat-spectrum sources - spectral 
coverage, which  ensures the great role of this instrument for
studies of GCRs  of intermediate 
energies, typically  between 1 GeV and 1 TeV. 
%
\begin{figure}[htbp]
\begin{center}
\includegraphics[width=0.7\linewidth]{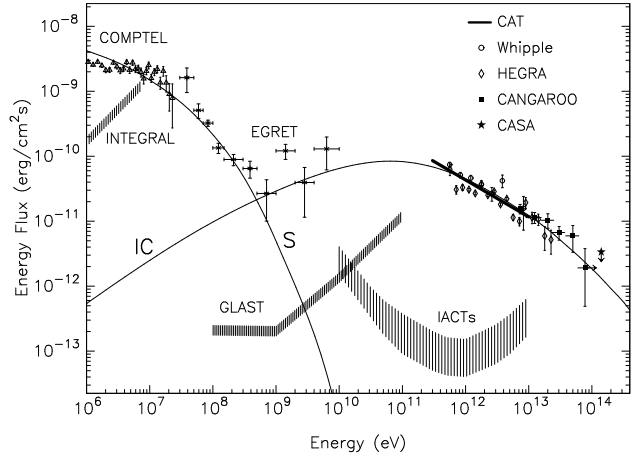}
\caption{The energy flux sensitivities of    
GLAST and  future IACT arrays. For comparison, the 
broad-band $\gamma$-ray spectrum of the Crab Nebula 
is also shown.} 
\end{center} 
\end{figure}
%
The energy  region of $\gamma$-rays beyond 100 GeV is the 
domain of ground-based gamma-ray astronomy, 
and could be best explored  by 
so-called  imaging atmospheric Cherenkov telescope (IACT)
systems (see e.g. \opencite{AhAk}).
The very high energy (VHE) region of $\gamma$-rays 
from 100 GeV to 100 TeV presents a special interest,
because these  $\gamma$-rays  carry direct information
about the sources responsible for formation of the 
most energetic part of the spectrum of GCRs 
extending to the so-called `knee' around $10^{15} \, \rm eV$.
The range of flux sensitivities that could be achieved by 
future IACT systems is shown in Fig.~1 \cite{GalTeV}. 
It is seen  that GLAST and future IACT arrays could probe 
point sources in a broad energy region from 0.1 GeV to 10 TeV 
at the energy flux level between $10^{-13}$ and  
$10^{-12} \, \rm erg/cm^2 s$. For comparison,  the flux of the Crab 
Nebula, the best studied 
galactic $\gamma$-ray source, 
exceeds $10^{-11} \, \rm erg/cm^2 s$ both at TeV and GeV 
energies (see Fig.~1).  
Thus, all extraterrestrial sources of high energy 
$\gamma$-rays with fluxes as low as 10 mCrab 
should be detected  by GLAST  and/or future IACT arrays.
In particular, these instruments should be able to 
reveal the sites  of most powerful accelerators 
of CR  protons in our Galaxy,  more specifically to detect 
$\pi^0$-decay $\gamma$-rays from  shell-type SNRs \cite{workinggroup}
as well as from other possible non-thermal galactic 
source populations like pulsars, X-ray binaries, 
microquasars, etc. (see e.g. \opencite{GalTeV}).   

In this paper I do not intend to review the 
high energy $\gamma$-ray sources, but rather I shall focus on 
a specific class of objects - giant molecular clouds (GMCs) 
located in the vicinity of CR accelerators. 
The existence of the CR  accelerator by itself is 
a necessary but not sufficient condition for effective 
$\gamma$-ray production; 
obviously we need the second component,  a {\em target}. 
The GMCs seem to be ideal objects  to play that role in our Galaxy. 
These objects are connected with star formation regions which are believed 
to be the most probable  (with or without SNRs) birth places of 
GCRs \cite{mont,paul}.  

\section{Giant Molecular Clouds as tracers of cosmic rays}

One of the principal parameters which determine the $\gamma$-ray 
visibility of GMCs is $M_5/d_{\rm kpc}^2$, 
where $M_5= M_{\rm cl}/10^5 M_{\odot}$ 
is the diffuse mass of the cloud located at a distance  
$d_{\rm kpc} = d/1 \, \rm kpc$ from the observer.
Assuming that $\gamma$-rays  are produced by
interaction of CRs with the ambient gas, 
\begin{equation}
F_{\gamma}(\geq E_\gamma) \simeq 10^{-7}  \left(\frac{M_5}
{d_{\rm kpc}^2}\right) q_{-25}(\geq E_\gamma)  \, \rm \; cm^{-2} s^{-1} \, ,
\end{equation}
where  $q_{-25}(\geq E_\gamma)=q_\gamma (\geq E_\gamma) 
/10^{-25} \, \rm (H-atom)^{-1} s^{-1}$ is the $\gamma$-ray emissivity.
In a ``passive'' GMC, i.e. in a cloud submerged in the 
``sea'' of GCRs, assuming that the level of the ``sea'' of GCRs 
is the same as the proton flux measured at the Earth, 
\begin{equation}
J_{\odot}^{\rm (p)}(E)=2.2 \, E_{\rm GeV}^{-2.75} 
\, \rm cm^{-2} s^{-1} sr^{-1} GeV^{-1} \, ,
\end{equation}
the $\gamma$-ray emissivity above 100 MeV
is equal to  
$q_{-25}(\geq 100 \, \rm MeV)=1.53  \ \eta 
\ \rm (H-atom)^{-1} s^{-1}$, where the parameter 
$\eta  \simeq 1.5$ takes into 
account the contribution of nuclei both in CRs and in the 
interstellar medium \cite{dermer}. A ``passive'' cloud can be detected 
at the EGRET sensitivity level,
if $M_5/d_{\rm kpc}^2 \geq 10$, taking into account 
large angular size (typically, several degree 
or more) of objects with $M_5/d_{\rm kpc}^2 \sim 10$.  
At energies $E \gg 1 \, \rm GeV$, 
$J(\geq E) \simeq 1.5 \times 10^{-13} \ 
(E/1 \, \rm TeV)^{-1.75} (M_5/d_{\rm kpc}^2) \, \rm ph/cm^2 s$ 
\cite{aharon91}, 
thus even for $M_5/d_{\rm kpc}^2 \sim 10$ 
(there are only several clouds in the Galaxy with such a large value), 
detection of $\gamma$-rays
from ``passive'' clouds 
above 10 GeV is an extremely difficult task both for GLAST and 
ground-based $\gamma$-ray detectors.  

Nevertheless, searches for very high energy $\gamma$-rays from GMCs present 
a certain  interest because of possible  existence of high-energy CR 
accelerators nearby or inside GMCs \cite{wolf,ormes,balaton,cloud}.
%
\begin{figure}[t]
\begin{center}
\includegraphics[width=0.45\linewidth]{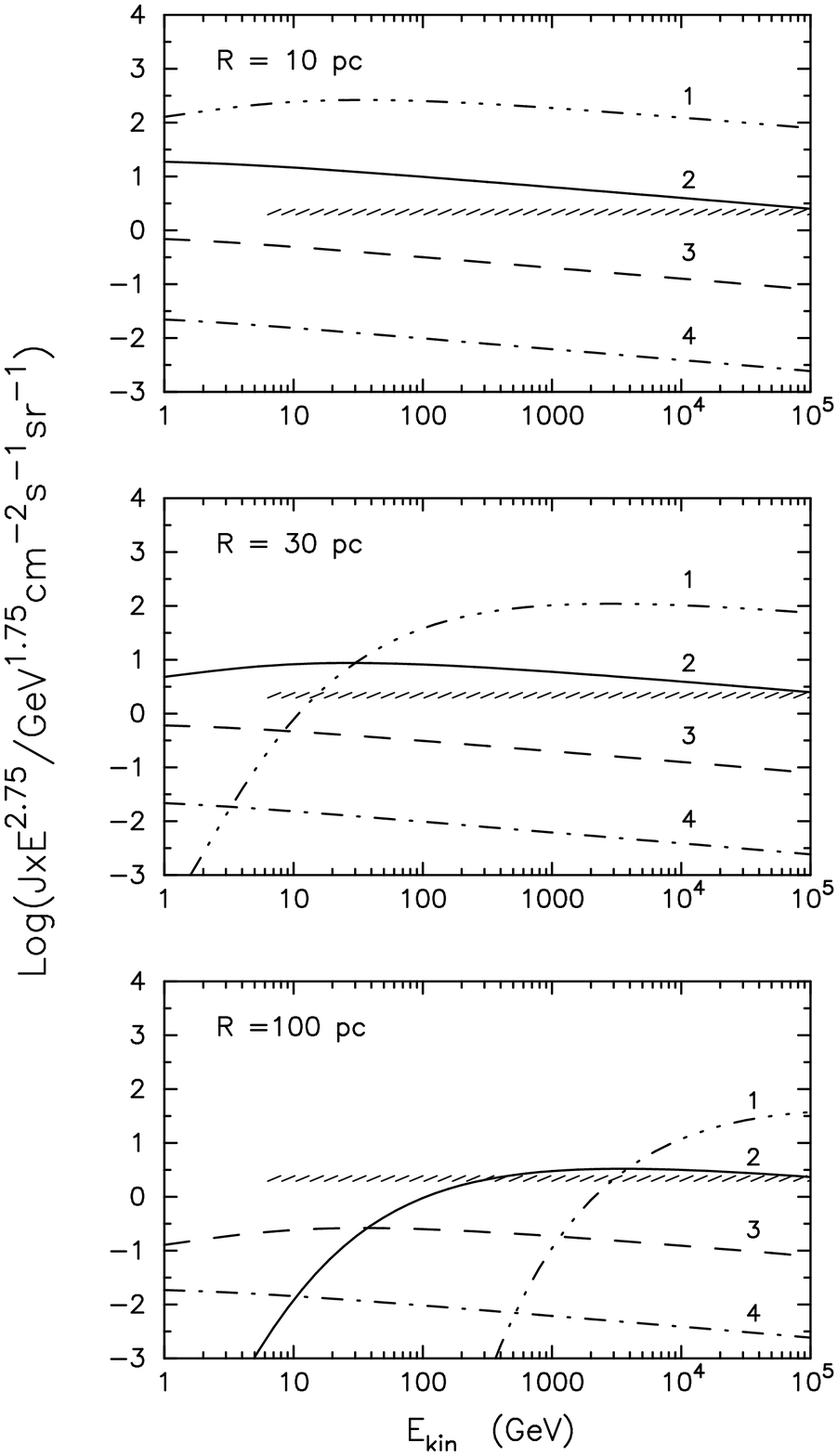}\hspace{5mm}
\includegraphics[width=0.45\linewidth]{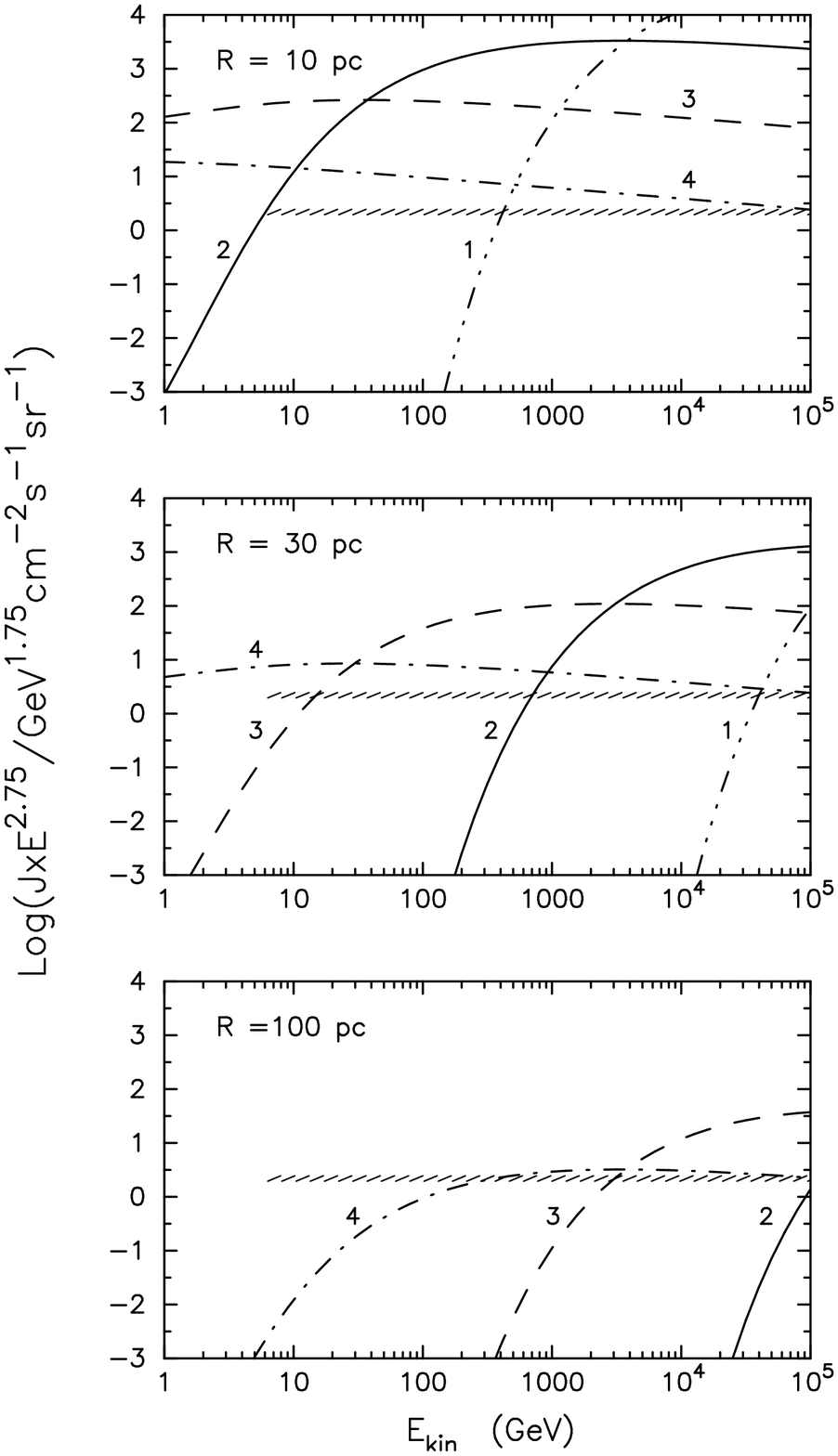}
\caption{Temporal and spectral evolution of CR fluxes 
at different (10 pc, 30 pc, and 100 pc) distances from
an {\it impulsive} proton accelerator. 
The power-law proton spectrum with 
$\alpha =2.2$ and total energy $W_{\rm p}=10^{50} \, \rm erg$
is supposed. The curves 
1, 2, 3, and 4  correspond to the
age of the source
$t=10^3\,\rm yr$, $10^4 \,\rm yr$, $10^5 \,\rm yr$, and $10^6\,\rm yr\,$,
respectively. The energy-dependent diffusion coefficient $D(E)$ with
power-law index $\delta =0.5$ is supposed. {\it Figure~2a} (on the left)
and {\it Figure~2b} (on the right) are for
$D_{10} =10^{28}$ and $D_{10} =10^{26} \,\rm cm^2/s$, respectively. 
The hatched curve shows the local (directly measured) flux 
of CR protons. (From Aharonian and Atoyan, 1996).}
\end{center}
\end{figure}
%
Indeed, assuming that CRs with total energy $W_{\rm p}$ injected by a
source into the ISM during their propagation
reach at the instant $t$ a radius $R(t)$, the mean
energy density of CRs is equal to
$w_{\rm p} \approx
0.5\, (W_{\rm p}/10^{50} \, \rm erg)(R/100
\, \rm pc)^{-3} \; \rm eV/cm^3$.
Thus, in the regions up to 100 pc around the CR
accelerators with $W_{\rm p} \sim 10^{50} \, \rm erg$,
the density of relativistic particles
at some stages, depending on the time history of
particle injection and the character of
their propagation in the ISM,
may significantly exceed the average level of
the ``sea'' of GCRs,  $w_{\rm GCR} \sim 1 \, \rm eV/cm^3$.
%
\begin{figure}[htbp]
\begin{center}
\includegraphics[width=0.65\linewidth]{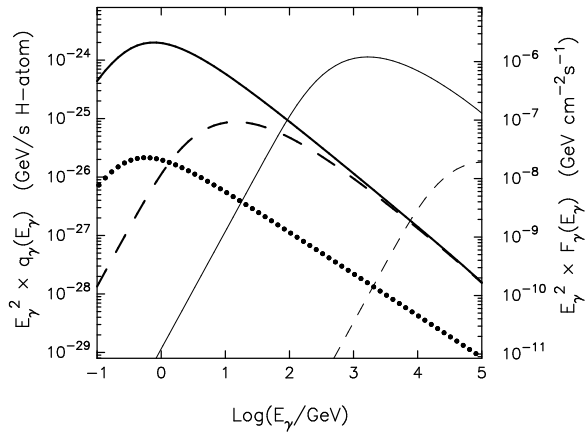}
\caption{Gamma-ray  emissivities in terms of 
$E_{\gamma}^2 \times q_\gamma(E_{\gamma})$  
at different times $t$ and distances $R$ from a 
proton accelerator.
The right-hand side axes shows the $\gamma$-ray
fluxes, $E_{\gamma}^2 \times F_\gamma(E_{\gamma})$,
which are expected from a cloud  with
parameter $M_5/d_{\rm kpc}^2 =1$.
The thin and heavy curves correspond to times $t=10^3$
and $10^5\,\rm yr$, respectively. 
The fluxes at distances $R=10$ and $30\,\rm pc$ 
are shown by solid and dashed lines.
The power-law index, the total energy of  
protons, and the diffusion coefficient 
are the same as in Fig.~2b. 
The curve shown by full dots corresponds to  the $\gamma$-ray 
emissivity (and flux) calculated for local CR protons. 
In order to take into account the contribution of nuclei, 
all curves should be increased by a factor of $\eta \approx 1.5$.
(From Aharonian and Atoyan, 1996).} 
\end{center} 
\end{figure}
%
In Fig.~2 the differential fluxes of protons, at distances $R=$10,
30, and 100 pc from an ``impulsive'' accelerator,  with total energy 
$W_{\rm p}=10^{50} \, \rm erg$ released in the form
of relativistic protons, are shown. The spectrum
of CRs at the given time and coordinates can significantly differ from 
the source spectrum. While the energy-independent diffusion 
leads to variation of the CR flux without change of the spectral
form, the energy-dependent diffusion results in
a modification of the particle spectrum. In Fig.~2
the diffusion coefficient is assumed in a power-law form,    
$D(E) \propto E^{0.5}$ above 10 GeV,
and constant below 10 GeV. The commonly used value for diffusion 
coefficient at 10 GeV, $D_{10}=D(10 \rm GeV)$,
is about  $10^{28} \, \rm cm^2/s$, 
however smaller values,  e.g.  $D_{10}=10^{26} \, \rm cm^2/s$  
cannot be excluded,  especially in the active star formation regions. 

The emissivity of the $\pi^0$-decay $\gamma$-rays within 30 pc  
of relatively young, $t \leq 10^5 \, \rm yr$, CR sources   
may significantly exceed the $\gamma$-ray emissivity introduced  
by the ``sea'' of GCRs, if the energy
release in relativistic particles  per accelerator, like a supernova 
remnant or a pulsar, $W_{50} \geq 0.01$ (see Fig.~3).   
Under these circumstances the existence of massive gas targets like 
molecular clouds in these regions may result in  
$\gamma$-ray  fluxes detectable by EGRET, if 
$(W_{50} \cdot M_5)/d_{\rm kpc}^2 \sim 0.1$.
In the case of energy-dependent propagation of CRs,
large variety of $\gamma$-ray spectra is expected,  
depending on the age of the accelerator, duration of injection, 
the diffusion coefficient, and the location of the cloud 
with respect to the accelerator. Therefore, 
at least part of the low-latitude unidentified  
EGRET sources,  exhibiting a broad 
range of photon indices, $\Gamma \sim 1.7-3.1$ 
\cite{merck},  could be explained as fortuitous location 
of CR accelerators in the proximity of massive molecular 
clouds. A special interest could present also 
the mid-latitude EGRET sources which likely associate with 
the Belt Gould of massive stars and gas clouds 
\cite{grenier,Gehrels}. If this is the case,
GLAST will be able to identify  these EGRET hot spots, as well as 
to  detect many more similar, but fainter 
(by one or two orders of  magnitude) objects.   

The comparison of $\gamma$-ray fluxes from clouds located at
different distances from an  accelerator,  may provide 
unique information about the CR diffusion coefficient $D(E)$. 
Similar  information may be obtained from a {\it single} 
$\gamma$-ray emitting cloud, but in different energy domains.
For the energy-dependent 
propagation of CRs  the probability for simultaneous detection of 
a cloud in GeV and TeV $\gamma$-rays
is  not very high, because the maximum fluxes at these energies
are reached at different epochs (see Fig.~3).
The higher energy particles
propagate faster and reach the cloud earlier, therefore
the maximum of GeV $\gamma$-radiation appears at the epoch when
the maximum of the TeV $\gamma$-ray flux is already over.
Meanwhile, in the case of energy-independent propagation (e.g.
due to strong convection) the ratio 
$F_\gamma(\geq 100 \, \rm MeV)/F_\gamma(\geq 100 \, \rm GeV)$ is independent
of time, therefore the clouds which are visible for EGRET 
at GeV energies would be detectable also at very high energies, 
provided that  the spectral index
of the accelerated protons  $\alpha \leq 2.3$.

\section{On the level of the ``sea'' of galactic cosmic rays}

The  effective mixture of CRs from individual sources, 
during their propagation in the interstellar magnetic fields 
on time-scales $\tau \geq 10^{7} \, \rm yr$, determines the average density 
of CRs throughout the galactic disk - the level of the ``sea'' 
of GCRs. The expected small variation of the  
latter on large galactic scales does not, however, exclude 
non-negligible gradients of the CR flux on smaller 
(typically $\le 100$ pc) scales, in particular in the proximity 
of young CR accelerators. 

It is generally believed that  the {\em local}, i.e. 
directly measured at the Earth, flux of CRs (LCRs) gives  
a correct estimate for the level of the ``sea'' of GCRs. However, 
strictly speaking,  this is an {\em ad hoc} assumption. In fact, 
it is not obvious that LCRs should be taken as undisputed 
representative of the whole galactic population of 
relativistic particles. In other words, we cannot exclude 
that the flux of LCRs could be dominated 
by a single or few local sources\footnote{This statement is certainly true 
for  the observed TeV electrons 
which suffer severe  synchrotron  and inverse Compton losses, 
and thus reach us, for any 
reasonable assumption about the diffusion coefficient, 
from regions no farther than a 
few hundred  parsecs \cite{Nishi,AAV}.}, especially given the fact that the 
Solar  system is located in a extraordinary region - 
inside active star formation complexes of 
the the so-called Gould Belt with essentially (more than  factor of 3) 
enhanced core-collapse supernovae rate \cite{grenier2}. 
Because of possible contamination of the 
``sea'' of GCRs by nearby CR sources, we may expect 
significant deviations of both  the spectrum and the energy density 
of LCRs from the spectrum and density of GCRs.
Apparently, the ``GCRs$\equiv$LCRs'' is an attractive  {\it hypothesis} 
which however still lacks robust {\it observational} tests. 
Such an inspection can be uniquely done by observations of  
high energy $\gamma$-rays from GMCs. 
Although this method generally 
requires detailed spectroscopic measurements 
for an ensemble of individual GMCs with known 
distances $d_{\rm kpc}$ and masses $M_5$, 
a reliable detection of 
$\gamma$-radiation from even a single {\em under-luminous} GMC,  
i.e. a cloud  emitting $\gamma$-rays below the  expected ``standard'' 
flux (dotted  curve in Fig.~3),
would imply that the CR flux we observe at 
the Earth is our local {\em ``fog''} obscuring the genuine flux of GCRs.
Below I argue that the $\gamma$-observations of the Orion complex  
contain evidence for such a dramatic  conclusion with important 
astrophysical implications for the origin of CRs.   
%
\begin{figure}[t]
\begin{center}
\includegraphics[width=0.485\linewidth]{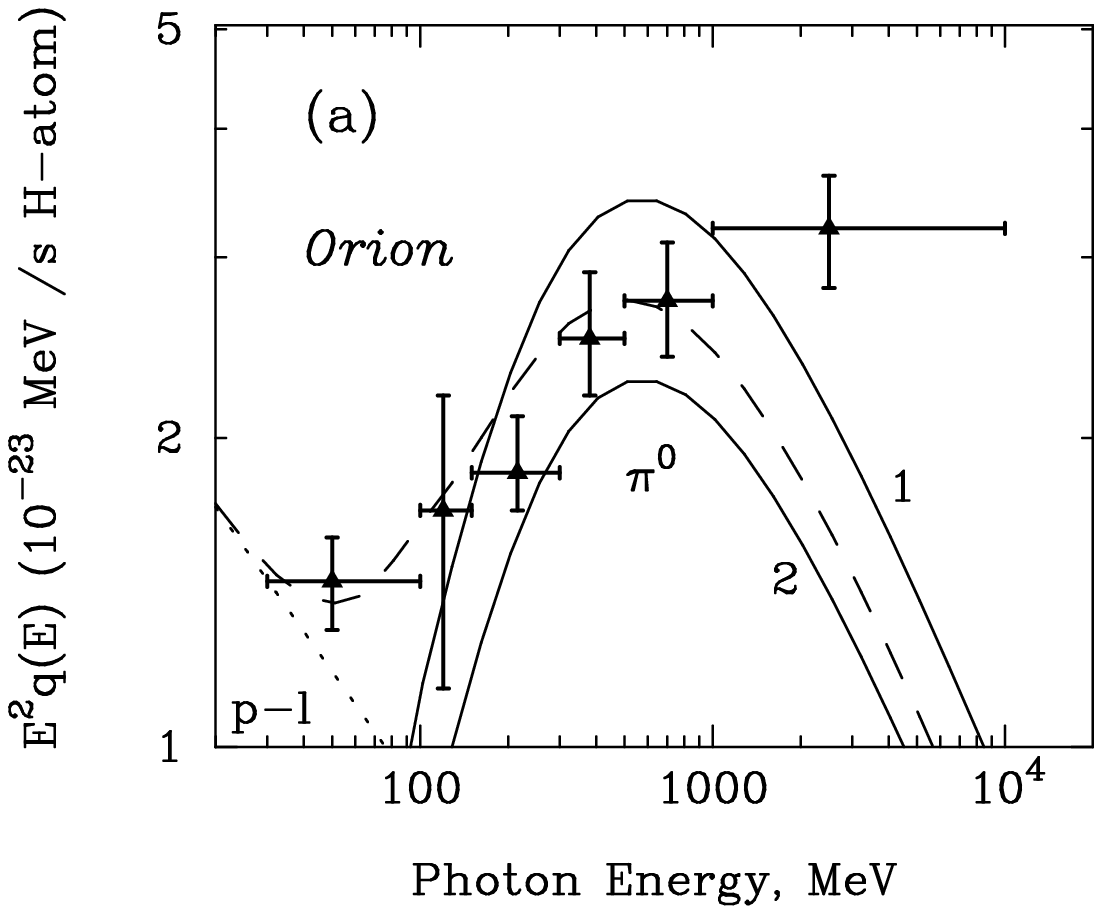}\hspace{1mm}
\includegraphics[width=0.42\linewidth]{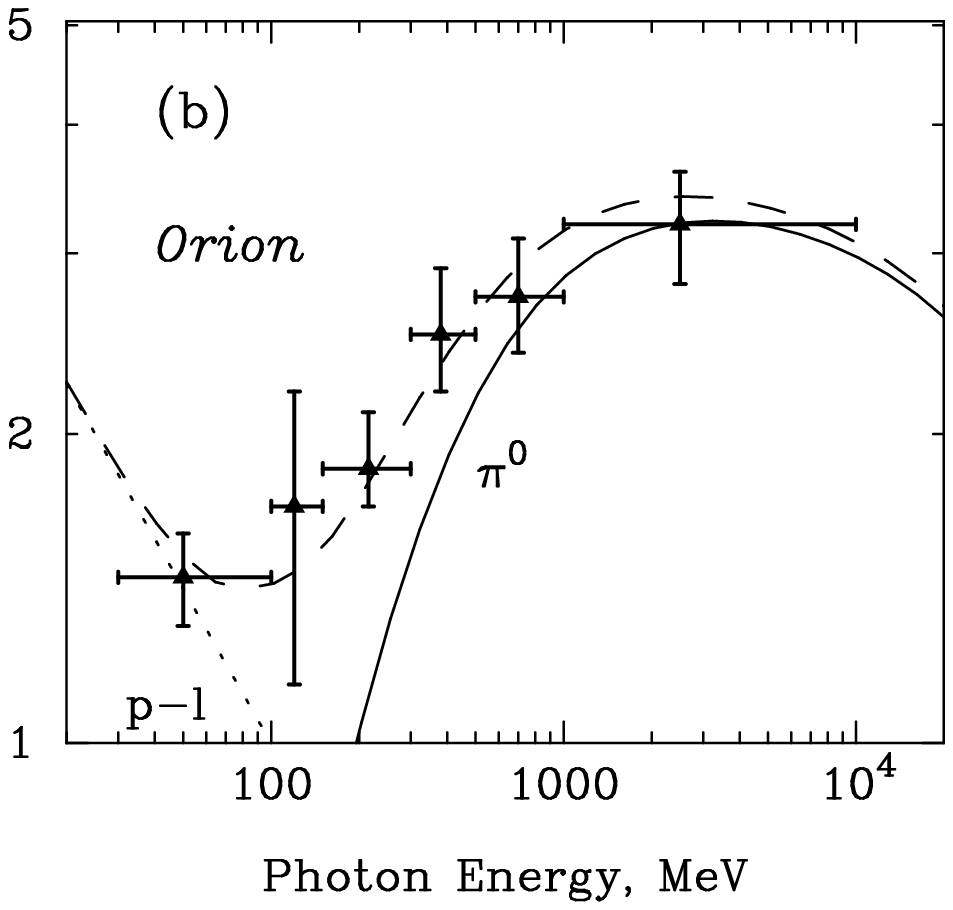}
\caption{Gamma-ray emissivity  
in the Orion region. The experimental points are from 
Digel et al. (1999).
The spectra of $\pi^0$-decay 
$\gamma$-rays 
are shown by solid lines. 
The power-law ({\em p-l}) components at low energies, with the photon index
$\Gamma=2.4$, are shown by dotted lines. 
The dashed lines correspond to the total, 
`` $\pi^0$+{\em p-l} '' spectra. {\bf (a)} $\pi^0$-decay 
$\gamma$-ray emissivity calculated for the LCRs flux (curve 1), 
and for a  1.5 times lower flux (curve 2).
The dashed curve corresponds to the sum of the latter and the 
low energy {\em p-l} component; {\bf (b)} $\pi^0$-decay 
$\gamma$-ray emissivity (solid curve) calculated for the 
CR proton flux  with power-low index index $\alpha=2.1$ and 
energy density  $w=0.7 \, \rm eV/cm^3$. The dashed curve corresponds 
to the superposition of the  $\pi^0$-decay 
and {\em p-l} components. Note that the flux of the {\it p-l} component 
is assumed somewhat higher than in Fig.~4a.} 
\end{center}
\end{figure}
%
\begin{figure}[htbp]
\begin{center}
\includegraphics[width=0.75\linewidth]{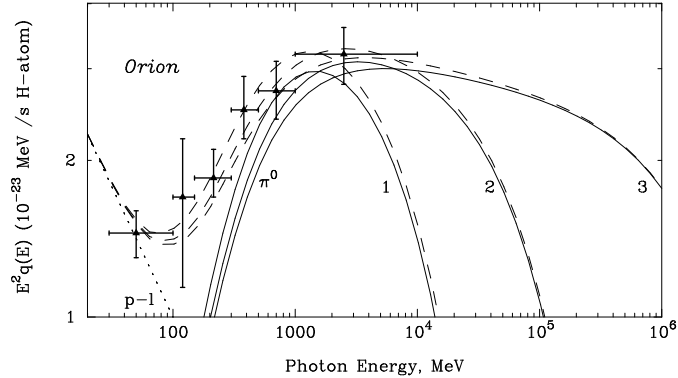}
\caption{Differential $\gamma$-ray emissivities in the 
Orion region. The curves 1, 2 and 3 correspond to 
emissivities calculated for CR spectra
with the same power-law  index $\alpha=2.1$,
but with exponential cutoff at 3 different energies,
$E_0=$100 GeV, 1 TeV, and 10 TeV, respectively. 
The corresponding energy densities of CR protons are:
$w_{\rm p}=0.55$, 0.7, and 0.85 $\rm eV/cm^3$. 
At low energies an additional power-law component of radiation
($\propto E^{-2.4}$)  is also assumed (dotted curve). The 
superpositions of $\pi^0$-decay
and power-law components are shown by dashed curves.}
\end{center} 
\end{figure}
%

The extensive study of high energy diffuse emission 
towards Orion  by EGRET allowed an accurate measurement 
of the  emissivity in the Orion region above 100 MeV,
$q_\gamma/4 \pi=(1.65 \pm 0.11) \times  10^{-26} \, \rm s^{-1} sr^{-1}$, 
and, more importantly, a derivation of the {\it  differential} emissivity 
in a broad, 30 MeV to 10  GeV energy region \cite{digel}. 
These measurements, together with $\gamma$-ray emissivities 
calculated under different  assumptions about the 
CR flux and spectrum, are shown in Fig.~4.  
The emissivity of  $\pi^0$-decay 
$\gamma$-rays that corresponds to the local proton flux represented by 
Eq.~(2), assuming an additional 50 per cent  contribution from 
$\alpha$-particles and other nuclei, 
is shown  in Fig.~4a (curve 1).  
Although the predicted  integral emissivity above 100 MeV,
$q_\gamma/4 \pi \simeq 1.8 \times  10^{-26} \, \rm s^{-1} sr^{-1}$,
almost coincides with the measured emissivity, it is difficult  
to find a good fit to the derived $\gamma$-ray spectrum without
assuming an additional low energy, e.g. a  power-law  
component (dotted line).  The latter  
with photon index $\Gamma \sim 2.3 - 2.5$ could be naturally 
attributed to the bremsstrahlung of primary or secondary
(i.e. $\pi^{\pm}$-decay) electrons.   
It is seen from Fig.~4a that the superposition of the `power-law' and  
`$\pi^0$-decay' contributions   
satisfactorily fits (dashed line) the derived experimental
points, if we ignore 
the last,  1 to 10 GeV EGRET point. 
Note, however,  that such a fit is achieved assuming  
a reduced, by a factor of 1.5,  CR flux in Orion compared 
with the local CR flux. Perhaps, we should not overemphasize 
this  difference, taking into account non-negligible
systematic errors in the reported fluxes of  
local CRs, as well as uncertainties in the  hydrogen column density 
used for derivation of the $\gamma$-ray emissivity.  

However, the discrepancy becomes really significant
if we include in the fit  the measured highest energy point  
at  1-10 GeV. This point  apparently requires harder CR spectrum. 
The lack of $\gamma$-ray measurements above 10 GeV do not
allow robust constraints on the power-low index, but, most probably,
it should not  exceed 2.3. In particular, Fig.~4b shows that 
a good fit to the EGRET data could be
achieved assuming a flat spectrum of protons with 
power-law index $\alpha=2.1$, and energy density
of particles below  $E \leq 1$~TeV at the level of 
$w=0.7 \, \rm eV/cm^3$,  
which is by a factor of 1.7 less than the energy density of 
local CR protons given by Eq.~(2).
Note that the fit in Fig.~4b does not include the $\gamma$-ray flux
produced by the ``sea'' of GCRs. This implies that  
the bulk of the observed $\gamma$-ray flux is contributed
by particles accelerated   
within  the Orion complex, with a significantly less 
(although unavoidable) contribution from GCRs  homogeneously 
distributed throughout the galactic disk, and freely passing 
through the molecular clouds.   
An extra contribution by GCRs around 1 GeV would lead to an overproduction of 
$\gamma$-rays, unless we assume that the level of the ``sea'' of GCRs  
at energies $\sim 1- 100 \, \rm GeV$ does not exceed,
for the given $\gamma$-ray flux uncertainties,   
approximately $1/3$ of the directly measured flux of LCRs. 
Such a  dramatic conclusion formally could be  
avoided  speculating  that CRs cannot freely 
penetrate into the dense clouds (see, however, 
\opencite{cesar}). For example, assuming that
the coefficient of reflectivity of CRs from a cloud 
decreases with energy, e.g. $\propto E^{-\kappa}$,
with $\kappa \sim 0.6-0.7$,  one can fix the required 
hard spectrum of particles without invoking  
additional CR accelerators inside Orion.  

Both these possibilities generally do not agree  with the 
conventional  models of galactic cosmic rays.  
Nevertheless, a thorough 
analysis based on future comprehensive $\gamma$-ray data is 
obviously needed 
before any strong claim about a necessity for revision of the 
current basic concepts of origin and propagation of CRs. 
In particular, it should be noticed that the conclusion 
about hard  CR spectrum in Orion is based merely on the  
reported $\gamma$-ray flux  above 1 GeV.
This indicates the importance of extension of accurate 
spectroscopic measurements of $\gamma$-radiation from Orion,
as well from other individual $\gamma$-ray emitting clouds  
detected towards Cepheus, Monoceros, and  Ophiuchus \cite{digel},
to the energy region beyond 10 GeV. Also, it would be 
important to investigate contributions of other 
$\gamma$-ray emission processes, in particular of the 
inverse Compton component contributed by relativistic electrons,
albeit in dense gas regions the $\pi^0$-decay radiation
is believed to be the  dominant $\gamma$-ray production channel.     

These measurements should allow an effective separation 
of contributions from  the ``sea'' of galactic CRs and from the   
local CR accelerators. It is especially important to extend 
the measurements to higher energies. If the ``excess'' 
flux at 1-10 GeV from Orion is indeed due to the  
local accelerator(s), we may expect continuation of this 
hard spectrum well beyond 10 GeV, depending on the 
high energy cutoff in the CR spectrum (see Fig.~5). On the other hand,
the hypothesis of the energy-dependent  reflection of galactic 
CRs from dense  clouds (assuming that GCRs$\equiv$LCRs), 
predicts an essentially different spectrum of $\gamma$-rays -    
a flat part below a few GeV, with a 
quite steep power-law tail  with  $\Gamma \simeq 2.75$,  at 
higher energies. This  part of the spectrum is  produced  
by protons of the ``sea'' of GCRs which freely
enter the cloud. Such standard high energy $\gamma$-ray 
tails should be observed from other local clouds as well.
     
Large-scale CO surveys of molecular clouds by 
Dame et al. (1987) revealed two dozens of local GMCs within 1 kpc 
with $M_5/d_{\rm kpc}^2 \geq 10$.  
The study  of high energy $\gamma$-rays from these clouds
with a broad-distribution of distances from $\sim 100-200 \,\rm pc$
(like the Taurus dark clouds or Aquila Rift) to 800 pc (like
Cyg OB7 and Cyg Rift), which should be easily visible for GLAST,
presents a great  interest for derivation of spatial and spectral 
distributions of CRs in our local environment. Moreover, it is likely
that GLAST will  be able to extent the study of massive GMCs, and therefore 
to probe the flux of galactic CRs, beyond 1 kpc.  



\end{article}
\end{document}